\documentclass[twocolumn,superscriptaddress,showpacs,amsmath,amssymb]{revtex4-1}

\usepackage{graphicx}
\usepackage{dcolumn}
\usepackage{bm}

\begin{document}

\preprint{APS/123-QED}

\title{Discontinuous Transition from a Real Bound State to \\ Virtual Bound State in a Mixed-Valence State of SmS}

\author{K. Imura}
 \altaffiliation{Present address: UVSOR Facility, Institute for Molecular Science, Okazaki 444-8585, Japan. E-mail address: imura@ims@ac.jp}
 \affiliation{Department of Physics, Graduate School of Science, Nagoya University, Nagoya 464-8602, Japan}
\author{S. Kanematsu}
 \affiliation{Department of Physics, Graduate School of Science, Nagoya University, Nagoya 464-8602, Japan}
\author{K. Matsubayashi}
 \affiliation{Institute for Solid State Physics, The University of Tokyo, Kashiwa 277-8581, Japan}
\author{H. S. Suzuki}
 \affiliation{Quantum Beam Center, National Institute for Materials Science, Tsukuba 305-0047, Japan}
\author{K. Deguchi}
 \affiliation{Department of Physics, Graduate School of Science, Nagoya University, Nagoya 464-8602, Japan} 
\author{N. K. Sato}
 \altaffiliation{E-mail address: kensho@cc.nagoya-u.ac.jp}
 \affiliation{Department of Physics, Graduate School of Science, Nagoya University, Nagoya 464-8602, Japan}

\date{\today}

\begin{abstract}
Golden SmS is a paramagnetic, mixed-valence system with a pseudogap. With increasing pressure across a critical pressure $P_{\rm c}$, the system undergoes a discontinuous transition into a metallic, anti-ferromagnetically ordered state. By using a combination of thermodynamic, transport, and magnetic measurements, we show that the pseudogap results from the formation of a local bound state with spin singlet. We further argue that the transition $P_{\rm c}$ is regarded as a transition from an insulating electron-hole gas to a Kondo metal, i.e., from a spatially bound state to a Kondo virtually bound state between 4$f$ and conduction electrons.
\end{abstract}

\maketitle

A class of materials called Kondo insulators (KIs) or semiconductors exhibit intriguing properties~\cite{Fisk}, such as a local bound state formation in SmB$_6$, YbB$_{12}$, and Sm$_{0.83}$Y$_{0.17}$S~\cite{Alekseev,Nemkovski,Alekseev2,Kasuya,Akbari}, which are ascribed to the interplay between 4$f$ and conduction electrons. In these mixed-valence compounds, Coulomb repulsion between 4$f$ electrons gives rise to a gap formation on the order 10 meV at the Fermi level. While such materials behave like a Kondo metal at temperatures ($T$) higher than the gap energy, they show insulating behavior at low $T$. For SmB$_6$, the collapse of the insulating gap and the emergence of a magnetic order occur at the same pressure~\cite{Barla2, Deer}, which suggests that the gap formation  competes with the long-range magnetic ordering.

Nonalloyed SmS undergoes a phase transition under pressure ($P$). With increasing $P$, the system undergoes a valence transition at a critical pressure of approximately $P =$ 0.7 GPa at $T =$ 300 K from divalence to mixed valence, accompanied by a color change from black to golden yellow~\cite{Wachter}. 
In this mixed-valence phase (named golden SmS), two configurations of 4$f^6$ and 4$f^5$ are energetically degenerate:
\begin{equation}
{\rm Sm}^{2+} (4f^6) \longleftrightarrow {\rm Sm}^{3+} (4f^5) + e.
\label{eq:vf}
\end{equation}
This is also described formally as ($1-\epsilon$) Sm$^{2+} + \epsilon$ Sm$^{3+}$ with 0 $\leq \epsilon \leq $1, which leads to the definition of the mean valence as $\nu=2(1-\epsilon)+3\epsilon$. Note that the Sm$^{2+}$ ion has no magnetic moment owing to the vanishment of total angular momentum. As seen from Fig.~1(a), $\nu$ estimated from high-$T$ experiments shows a continuous evolution with $P$~\cite{Coey, Annese, Deen}. A similar smooth $P$-variation is observed in the conductivity $\sigma$ measured at $T=$ 150 K [Fig.~1(b)]. From these results, it is inferred that the number of electrons $e$ on the right-hand side of eq.~(1) monotonically increases with $P$. As seen from the phase diagram of Fig.~1(c), there is no phase transition as a function of $P$ at high $T$.

\begin{figure}[b]
\begin{center}
\includegraphics[width=0.3\textwidth]{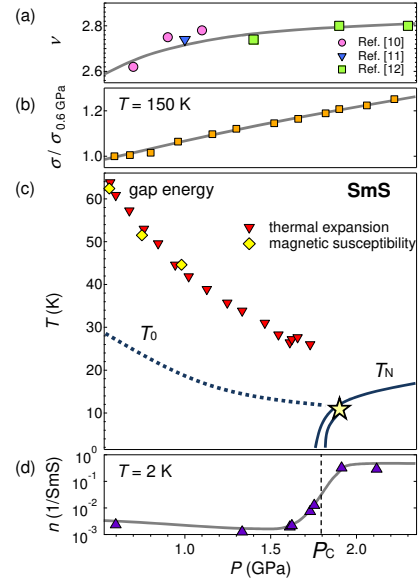}
\end{center}
\caption{(color online). Continuous pressure evolution of (a) mean valence $\nu$ at high temperature~\cite{Coey, Deen, Annese} and (b) conductivity ratio $\sigma(P)/\sigma(P=0.6$ GPa) at 150 K. For the definition of $\nu$, see text. (c) $P$-$T$ phase diagram of golden SmS. $T_0$ and $T_{\rm N}$ denote the crossover of a pseudogap opening and the N\'{e}el temperature, respectively~\cite{Matsubayashi1,Imura1}. The star marks the tricritical point that separates the first-order transition (double line) from the second-order transition (single line)~\cite{Imura1}. Triangles and diamonds show the charge gap $\Delta$ and spin gap energy $\Delta_{\rm M}$, respectively. (d) Steep pressure evolution of free-carrier number $n$ at $T=2$ K. Lines displayed in (a)-(d) are guides to the eyes .}
\label{f1}
\end{figure}

At low $T$, by contrast, a magnetic transition occurs at $P_{\rm c} \sim$ 1.8 GPa~\cite{Barla, Haga}. Magnetic susceptibility revealed that the ground state at $P>P_{\rm c}$ is antiferromagnetic (AFM)~\cite{Matsubayashi3}. This quantum phase transition (QPT) at $P_{\rm c}$ is of the first order as suggested by the large variation in the carrier number $n$ at $T =$ 2 K, as shown in Fig.~1(d), in accordance with a jump in the volume, as reported previously~\cite{Imura1}. The discontinuous transition denoted by the double line in Fig.~1(c) changes to a continuous transition at the tricritical point (TCP), as denoted by the single line and star, respectively. The broken curve in the paramagnetic phase, which terminates at the TCP, is a crossover line signifying the opening of a pseudogap at the characteristic temperature $T_0$~\cite{Matsubayashi1}.

Besides the extensive studies mentioned above, including a theoretical analysis, the origin of the pseudogap remains unknown. In this Letter, we present systematic experiments on golden SmS. The obtained results are summarized as follows. (i) All physical quantities studied here showed an anomaly at the same temperature $T_0$. (ii) In particular, the magnetic susceptibility exhibited spin-singlet state formation below $T_0$. To explain these results, we propose a scenario of exciton formation (i.e., the electron $e$ in eq. (1) being trapped at the Sm$^{3+}$ ion) below $T_0$, as theoretically predicted by Kikoin~\cite{Kikoin}.

Single-crystal preparation was presented in ref. 19. Experimental details of transport, thermal expansion, and thermoelectric power were described elsewhere~\cite{Imura2,Imura3}. DC magnetization was measured by a commercial SQUID magnetometer. High pressure was generated using a piston-cylinder clamped cell with Daphne oil 7373 as the pressure-transmitting medium. Applied pressure was estimated at low temperatures.

\vspace{5pt}
\noindent
{\it Low-temperature region at $P<P_{\rm c}$ ---} 
We start with the presentation of low-$T$ experimental results at $P<P_{\rm c}$. Figure 2(a) shows the temperature dependence of the electrical resistivity $\rho(T)$ of a Sm-excess sample. $\rho(T)$ increases with decreasing temperature and exhibits a hump structure, before increasing again below 10 K. Note that a similar structure was observed in all the other samples studied here with different compositions [see Figs.~2(b) and (c)] and in the samples presented in previous reports~\cite{Lapierre, Konczykowski}. These suggest that the anomaly can be intrinsic to golden SmS.

To explore the origin of the anomalous behavior, we measured several physical quantities at almost the same pressure ($P =$ 0.6 and 1.6 GPa). The results are shown in Figs.~3(a)-3(f).

\begin{figure}[h]
\begin{center}
\includegraphics[width=0.45\textwidth]{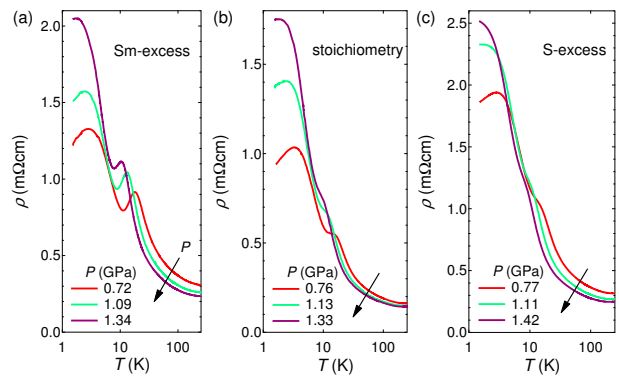}
\end{center}
\caption{(color online). Temperature dependence of electrical resistivity of (a) Sm-excess, (b) stoichiometric, and (c) S-excess crystals of SmS at pressures identified in the figures.}
\label{f2}
\end{figure}

\begin{figure}[h]
\begin{center}
\includegraphics[width=0.35\textwidth]{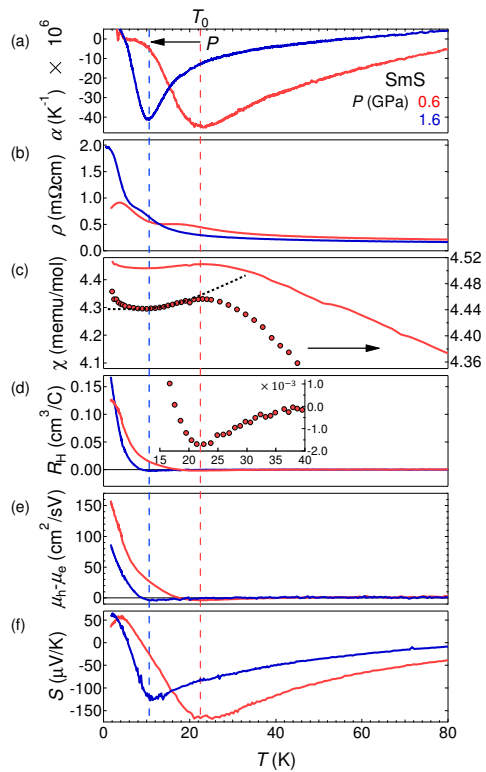}
\end{center}
\caption{(color online). Temperature dependence of physical quantities of golden SmS measured at $P=$ 0.6 and 1.6 GPa. (a) Thermal expansion coefficient $\alpha$, (b) electrical resistivity $\rho$, (c) magnetic susceptibility $\chi$, (d) Hall coefficient $R_{\rm H}$, (e) relative charge mobility $R_{\rm H}/\rho = \mu_{\rm h}-\mu_{\rm e}$, and (f) Seebeck coefficient $S$. Broken (vertical) lines denote $T_0$.}
\label{f3}
\end{figure}

(a) At 0.6 GPa, thermal expansion coefficient $\alpha(T)$ shows a broad negative minimum at $T_0 =$ 22 K. By subtracting the ambient result as a phonon contribution, we observed a Schottky-like anomaly with a (pseudo)gap $\Delta =$ 61 K~\cite{Matsubayashi1}. For the $P$ dependence of $\Delta$, see Fig.~1(c), in which one may notice that the gap disappears suddenly in the vicinity of $P_{\rm c}$.

(b) The temperature at which $\rho(T)$ shows the hump structure almost coincides with $T_0$ probed by the thermodynamic technique mentioned above. Therefore, the low-$T$ increase (below $T_0$) in the resistivity is clearly intrinsic and of bulk origin.

(c) According to a previous report~\cite{Wachter}, $\chi(T)$ monotonically increases with decreasing $T$, and saturates at a constant value at low $T$. A new finding in this study is a small but clear decrease below $T_0$ described as
\begin{equation}
\delta \chi(T) \propto \exp(-\Delta_{\rm M}/k_{\rm B}T)
\end{equation}
(see dotted line in the inset). The magnitude of the spin gap is estimated as $\Delta_{\rm M} = 62$ K at 0.6 GPa, which is almost the same as the pseudogap $\Delta$ obtained above.
By analogy to the local bound state in KIs, we interpret the spin gap as a signature of the local bound state formation with a spin singlet.

(d) The Hall constant $R_{\rm H}(T)$ exhibits a minimum at $T_0$ (see the inset). The major carrier at low $T$ is a hole, in contrast to an electron in the AFM ordered state~\cite{Imura4}. 
Assuming that $n_{\rm h} = n_{\rm e} = n$ for a compensated metal (where $n_{\rm h}$ and $n_{\rm e}$ denote the hole and electron carrier densities, respectively), we have the relation $R_{\rm H}=\frac{1}{nec}\frac{\mu_{\rm h}-\mu_{\rm e}}{\mu_{\rm h}+\mu_{\rm e}}$ (where $\mu_{\rm h}$ and $\mu_{\rm e}$ represent the hole and electron mobilities, respectively). This leads to the inequality $R_{\rm H} \leq (nec)^{-1}$ because $ |\frac{\mu_{\rm h}-\mu_{\rm e}}{\mu_{\rm h}+\mu_{\rm e}}| \leq 1$. Finally, we obtain the carrier number $n <$ 3 $\times$ 10$^{-3}$ per formula unit (f.u.)~at 2 K. This upper value of $n$ is two orders of magnitude smaller than the room-temperature value, indicating that the carrier density decreases due to the pseudogap opening below $T_0$.

(e) The ratio $R_{\rm H}/\rho$ yields the relative charge mobility $\mu_{\rm h} - \mu_{\rm e}$. Whereas $\mu_{\rm h} \sim \mu_{\rm e}$ at high $T$, $\mu_{\rm h} \gg \mu_{\rm e}$ at low $T$. This suggests that electrons become less mobile than holes at low $T$. Note that the mobility shows a strong $T$ dependence at low $T$. If the impurity scattering is dominant, then the mobility is $T$-independent. Therefore, the low-$T$ scattering mechanism is not dominated by impurity scattering but presumably by carrier-carrier scattering. This interpretation is consistent with the result derived from (d); the carrier density decreases with decreasing $T$.

(f) The Seebeck coefficient $S(T)$ shows a very similar feature to $\alpha(T)$. According to a theory of KI~\cite{Saso}, the peak of $|S(T)|$ indicates (pseudo)gap formation. The sign of $S(T)$ is indeed consistent with that of $R_{\rm H}(T)$.

\vspace{5pt}
\noindent
{\it Low-temperature region at $P>P_{\rm c}$ ---} 
In the AFM region above $P_{\rm c}$, the major carrier is an electron in contrast to that in the pseudogap region, and its density is approximately 0.5 per f.u.~[Fig.~1(d)]. This value is of the same order as that expected from the high-$T$ $\nu$ value [Fig.~1(a)], indicating that the carrier number is almost invariant with $T$.

Figure 4(a) shows $\rho(T)$ in a semi-logarithmic plot. We observe the resistivity to logarithmically increase with decreasing $T$ before decreasing at $T_{\rm N}$. At low $T$ [see Fig.~4(b)], $\rho(T)$ shows a Fermi liquid behavior, $\rho(T)=\rho_0 + A T^2$. The $A$-coefficient decreases with increasing $P$ away from $P_{\rm c}$, consistent with a previous report~\cite{Ohashi}. Note that the $A$-coefficient amounts to a large $A \sim 0.7$ $\mu \Omega$cm/K$^2$ at $P \simeq 2.2$ GPa, comparable to that of the Kondo metal CeB$_6$ in the AFM ordered state~\cite{Sato}. It is inferred from the so-called Kadowaki-Woods relation that the electronic specific heat coefficient of the present system exceeds 200 mJ/K$^2$. Therefore, golden SmS (at $P>P_{\rm c}$) is identified as a metallic heavy-fermion system, in which a virtual bound state forms at low $T$.

\begin{figure}[t]
\begin{center}
\includegraphics[width=0.45\textwidth]{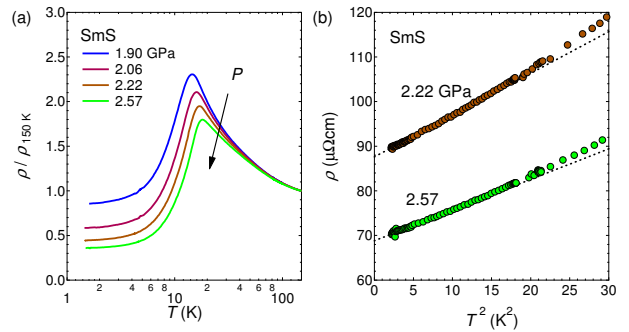}
\end{center}
\caption{(color online). (a) Temperature dependence of electrical resistivity ratio $\rho(T)/\rho(T=150 \rm K)$ at pressures identified in figure. (b) $T^2$ dependence of resistivity at $P=$ 2.22 and 2.57 GPa. Dotted lines denote the $T^2$ dependence.}
\label{f4}
\end{figure}

\begin{figure}[h]
\begin{center}
\includegraphics[width=0.45\textwidth]{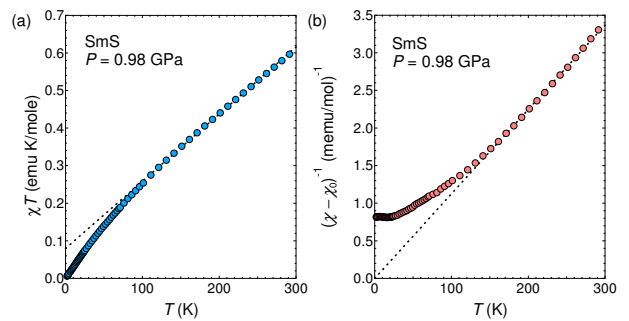}
\end{center}
\caption{(color online). Temperature dependence of (a) magnetic susceptibility $\chi$ multiplied by $T$ at $P=$ 0.98 GPa and (b) inverse of $\chi-\chi_0$. Here, $\chi_0$ is the $T$-independent term of $\chi(T)$. Dotted lines indicate $\chi(T)$ of eq.~(3) in the text.}
\label{f5}
\end{figure}

\vspace{5pt}
\noindent

{\it High-temperature region ---}
There is no phase transition at high $T$ over the entire $P$ region investigated here, as mentioned above. Let us consider $\chi(T)$ again. Figure~5(a) shows the $T$ dependence of the magnetic susceptibility in the form of $\chi T$ against $T$ at $P =$ 0.98 GPa. We observe a $T$-linear dependence in the high-$T$ region. This result allows us to describe the magnetic susceptibility as
\begin{equation}
\chi(T) \simeq  C/T + \chi_0, 
\end{equation}
with a $T$-independent term, $\chi_0 \simeq 1.74 \times 10^{-3}$ emu/mole f.u. This is confirmed from the plot of the data in the form of $(\chi- \chi_0)^{-1}$ against $T$ in Fig.~5(b).
Note that the Curie constant $C \simeq$ 0.089 emu K/mole f.u. is comparable to the free Sm$^{3+}$ ion value. In conjunction with the logarithmic $T$-dependence of $\rho(T)$ mentioned in Fig.~4(a), we suggest the presence of localized magnetic moments in the high-$T$ region.

\vspace{5pt}
Next, let us interpret the results presented above in terms of the exciton model~\cite{Kikoin}, assuming that exciton binding energy is given by $k_{\rm B}T_0$ (where $k_{\rm B}$ is a Boltzmann constant). At high $T$ greater than $T_0$, the bound state breaks down; hence, the Sm$^{3+}$ ion and the electron $e$ in eq.~(1) are decoupled and contribute to the Curie and $\chi_0$ terms in eq.~(3), respectively. Indeed, the electron $e$ carries an electrical current. Then, the interplay between the Sm$^{3+}$ ion and the electron $e$ can give rise to Kondo scattering, as shown in Fig.~4(a). At low $T$ (less than $T_0$), the binding energy overcomes the thermal energy, then the electron $e$ is trapped to form the bound state, resulting in the carrier number decrease shown in Fig.~3. Therefore, we conclude that the pseudogap results from exciton formation. Note that this is compatible with the absence of a hybridization gap in calculated band structures~\cite{Balle,Schumann,Harima}.

Finally, we attempt to estimate the exciton radius $a_{\rm B}$, assuming the well-known formula $a_{\rm B} = (E_{\rm H}/\epsilon \Delta)a_{\rm H}$, where $E_{\rm H}$ and $a_{\rm H}$ are the binding energy and Bohr radius of the hydrogen atom, respectively. Provided that the dielectric constant $\epsilon \sim$ 230~\cite{Travaglini}, which is obtained for a scraped surface at ambient pressure, and $\Delta \sim 26$ K at $P_{\rm c}$ [Fig.~1(c)], we have $a_{\rm B} \sim 1.4$ nm, which should be regarded as an upper limit of $a_{\rm B}$ because $\epsilon$ is expected to become greater with increasing $P$. Alternatively, we evaluate $a_{\rm B} \sim 0.2$ nm using the equation $a_{\rm B} = \lambda^{-1}$ (where $\lambda^{-1}$ is the screening length). Here, we assumed that the exciton breaks down due to the screening effect at $P_{\rm c}$, and estimated $\lambda$ by the Thomas-Fermi approximation $\lambda^{2} = 4n^{1/3}(3/\pi )^{1/3}/a_{\rm H}$ with $n = 1 \times 10^{-3}$ per f.u.~[see Fig.~1(d)]. These two values of $a_{\rm B}$ may not be incompatible with each other considering the crude approximations made. If we accept this, then we suggest that the screening effect plays a role in the phase transition at $P_{\rm c}$.

One may ask if excitons condense at low $T$. All of the physical properties studied here exhibit an anomaly at $T_0$, but there is no indication of the phase transition there. At the present stage, therefore, it is reasonable to assume that the ground state is not a Bose-Einstein condensate but an exciton gas. It may be interesting to explore if the condensation would occur at an extremely low $T$.

In summary, we have presented the thermodynamic, transport, and magnetic properties of golden SmS, and proposed an exciton model to interpret them. At high $T$, there are localized moments as confirmed from the Curie law. The ground state depends on pressure. At $P<P_{\rm c}$, the localized moments disappear as a result of the spin-singlet, exciton-like bound state formation below $T_0$. At $P>P_{\rm c}$, by contrast, the localized moments survive and form the AFM ordering below $T_{\rm N}$ owing to the Ruderman-Kittel-Kasuya-Yosida interaction via heavy quasiparticles. This means that the ground state changes from the real bound state to the Kondo virtual bound state at $P_{\rm c}$, which may result from the competing interactions of pseudogap formation and antiferromagnetic ordering under the influence of screening. These results demonstrate that golden SmS can be a model system in which the discontinuous QPT occurs between the valence fluctuation and Kondo metal regimes.

\section*{Acknowledgments}
The authors thank I. Jarrige, H. Yamaoka, S. Kimura, T. Ito, T. Saso, T. Kasuya, and J. Flouquet for fruitful comments.
This work was partially supported by KAKENHI (S) (No. 20224015). KI was supported by a Grant-in-Aid for JSPS Fellows.

\end{document}